# Study on the energy criterion of cuprate superconductivity

Gu Jiapu *

(GIRE Technology Co.Ltd., Chongqing, China, 400039)

## Abstract

In this paper, we use the variation of spontaneous magnetization to describe the influence of electron holes in cuprate superconductors, and use competitive energy relations to explore the superconductivity rule and energy criterion, on this basis, we can deduce a clear physical image of superconducting phase diagram and superconducting mechanism.

## 1. Introduction

The high-temperature superconducting mechanism is still studied currently, multitudinous theories can only explain partial experiment phenomenon, cannot explain the phase diagram completely. the main difference of current theories is the formation mechanism of electron pairing, a major cause of disputation is the microstructure complexity and the corresponding mathematical complexity.

In this paper, we try to evade the complexity as indicated above, we use the variation of spontaneous magnetization to describe the influence of electron holes, and use competitive energy relations to explore the superconductivity rule, on this basis, we can deduce a clear physical image of superconducting phase diagram.

## 2. The theory foundation

This paper is based on the following views by experiment support:

（1）Two electron states structure is considered reasonable, namely the Cu3d electrons constitute antiferromagnetic background, the O2p orbital electron holes itinerate and electron pairs form in certain conditions.

(2) The interaction between electron holes and antiferromagnetic background is favorable for forming superconductivity, but the increment of electron holes will destroy further the antiferromagnetic background, this is the dual character of electron holes.

(3) Magnetic fluctuation is considered a major cause of electron pairs[1,2,3], but electron-phonon interaction cannot be neglected in the meantime[4,5,6].

---

*Corresponding author: Email: gujiapu@yahoo.com.cn, Tel: (86)13108965951

In this paper, we think the above-mentioned contents are the pivotal basis of superconductivity theory construction. If a mathematics expression can clearly express above contents, on this basis we can understand and deduce other intricate phenomena such as pseudogap, quantum critical point 0.19, etc.

## 3. The mathematical description of magnetic fluctuation energy

In this paper, we think the magnetic fluctuation energy is the main source of electron pairing energy.

At first, we use the variation of spontaneous magnetization to describe the influence of electron holes. to the undoped parent compound of CuO2 surface, suppose the spontaneous magnetization parameter M for sublattice , by molecular field approximation, the corresponding magnetic exchange energy $E_0$:

$$E_0 = \frac{1}{2}\lambda M^2,$$

By the doped concentration x of O2p orbital electron holes, the spontaneous magnetization is weakened, at the moment the spontaneous magnetization is thought to turn to (1-x)M linearly, the corresponding magnetic exchange energy $E$:

$$E = \frac{1}{2}\lambda(1-x)^2 M^2.$$

The variation of magnetic exchange energy $\Delta E = E_0 - E$ is as follows:

$$\Delta E = \frac{1}{2}\left[1-(1-x)^2\right]\lambda M^2.$$

the variation energy $\Delta E$ namely the magnetic fluctuation energy has net attractive potential for ambient electrons, with the magnetic steering of antiferromagnetic background, two adjacent cruising electrons can be turned into cooper pairs easily. so the doped hole of O2p orbital can be considered the attraction medium of cooper pairs. but the attractive potential characteristic of $\Delta E$ can be also a competing factor to farther disturb and damage the antiferromagnetic background, this is the dual character of the magnetic exchange fluctuation energy $\Delta E$.

## 4. The energy relations in superconductive state（omitting electron-phonon interaction）

When we omit the electron-phonon interaction temporarily, the magnetic fluctuation energy $\Delta E$ can be regarded as the only source of electron pairing energy. Between the medium hole and one pairing electron, the interrelated competing parameters have the coulomb potential energy V, the thermal vibration energy $k_B T$, the weakened magnetic exchange energy $\frac{1}{2}\lambda(1-x)^2 M^2$, their competition relations are as follows:

$$\Delta E \geq V + k_B T \qquad (4\text{-}1)$$

$$\Delta E + V + k_B T \leq \frac{1}{2}\lambda(1-x)^2 M^2 \qquad (4\text{-}2)$$

The above inequality (4-1) describes the pairing attraction characteristic of the magnetic fluctuation energy $\Delta E$. in the superconducting state, the attraction energy $\Delta E$ can resist the destructive influences of coulomb repulsion and thermal vibration.

The above inequality (4-2) describes the farther destructive influence of the magnetic fluctuation energy $\Delta E$ on the weakened magnetic exchange energy. In the meantime, the coulomb potential energy V and the thermal vibration energy $k_B T$ have the same farther destructive influence on the weakened magnetic exchange energy. But in the superconducting state, it is precondition that the destructive influences cannot destroy the weakened magnetic exchange energy namely the antiferromagnetic background.

The above inequalities (4-1, 4-2) can be simplified to T~x formats as follows:

$$T \leq -\frac{1}{2k_B}\lambda M^2(1-x)^2 + \frac{1}{k_B}(\frac{1}{2}\lambda M^2 - V) \qquad (4\text{-}3)$$

$$T \leq \frac{1}{k_B}\lambda M^2(1-x)^2 - \frac{1}{k_B}(\frac{1}{2}\lambda M^2 + V) \qquad (4\text{-}4)$$

$$(0 \leq x \leq 1, \quad T \geq 0)$$

Now we establish the "T—x" rectangular coordinate system, under the constraint conditions "$0 \leq x \leq 1, \ T \geq 0$", the both conditions hold of inequations (4-3, 4-4) is just the inclined shadow area "AHB" below, namely the superconductivity area:

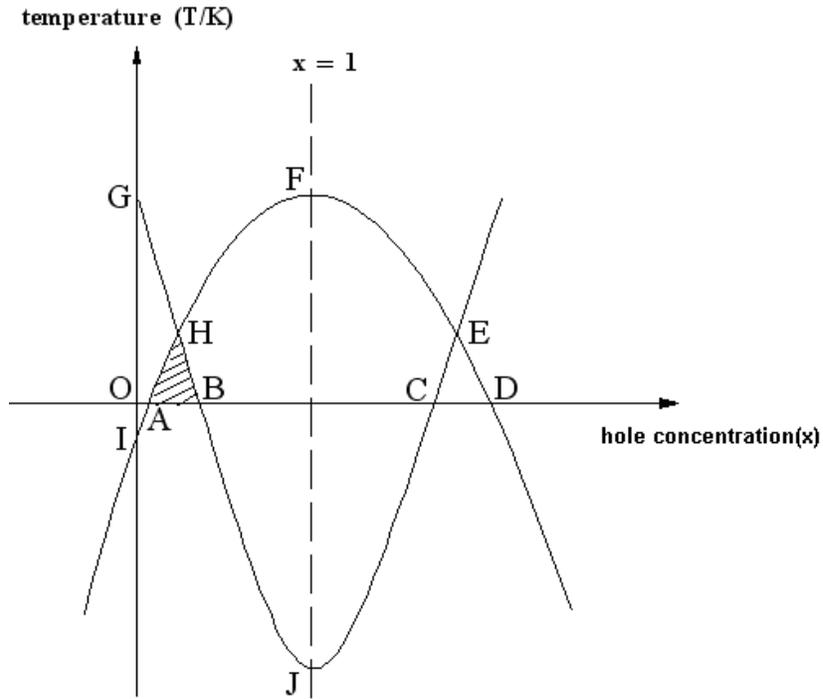

**Fig.1:** The "T-x" relation in superconductive state（omitting electron-phonon interaction）.

Moreover, when we remove the $(1-x)^2$ part of simultaneous inequations (4-3, 4-4), we get the relational expression as follow:

$$\frac{1}{2}\lambda M^2 \geq 3V + 3k_B T \qquad (4\text{-}5)$$

It is the precondition of superconductivity energy, having nothing to do with the electron hole concentration.

## 5. The characteristics derivation of cuprate superconductivity by magnetic fluctuation action

Now we use the inclined shadow area AHB to study the cuprate superconductivity.

In Fig.1，the coordinate values of point A, B, H are as follows：

$$A: (1-\sqrt{1-\frac{2V}{\lambda M^2}}, 0)$$

$$B: (1-\sqrt{0.5+\frac{V}{\lambda M^2}}, 0)$$

$$H: (1-\sqrt{\frac{2}{3}}, \frac{\lambda M^2}{6k_B}-\frac{V}{k_B})$$

From the superconductivity area AHB, we can deduce the important superconductivity phenomena as follows:

(1) Existing the weak doping area and over-doping area in cuprate superconductivity state, they respectively are corresponding to the curved section AH, HB in Fig.1.

(2) The low concentration characteristics of doping holes and current carriers:

When $\frac{1}{2}\lambda M^2 \gg 3V$, we get the ultimate concentration:

$$x_{A\min} = 0,$$
$$x_{B\max} = 1-\sqrt{0.5} \approx 0.2929$$

Namely the hole concentration has the following limit range in superconducting state:

$$x \in (0, \ 0.2929)$$

so we think the doping holes and current carriers have low concentration characteristics.

(3) Existing pseudogap in weak doping area, it is corresponding to the curved section GH. the pseudogap curved section GH and the superconducting energy gap HB in over-doping area are two parts of the same smooth curve GB, so we can think that they have a same physical origin.

(4) With the increasing of hole concentration, the superconducting energy gap AH in weak doping area becomes the non- superconducting energy gap HF part finally, we think they also have another same physical origin.

(5) Existing the quantum critical point H [7], it is the optimum doping point having the highest superconducting transition temperature value (omitting the effect of electron-phonon interaction), and it is corresponding to a constant value of hole concentration $x_H = 1-\sqrt{\frac{2}{3}} \approx 0.1835$.

# 6. The influence of electron-phonon interaction on real superconductivity phase diagram

We think the electron-phonon interaction energy $W$ between the medium hole and one pairing electron has a dual characteristic as the magnetic fluctuation energy $\Delta E$, they are parallel relation, so we use the ($\Delta E + W$) to replace the parameter $\Delta E$ in ineq.(4-1, 4-2) as following:

$$\Delta E + W \geq V + k_B T \qquad (6\text{-}1)$$

$$\Delta E + W + V + k_B T \leq \frac{1}{2}\lambda(1-x)^2 M^2 \qquad (6\text{-}2)$$

To farther rewrite the above inequalities into the following T~x forms:

$$T \leq -\frac{1}{2k_B}\lambda M^2(1-x)^2 + \frac{1}{k_B}(\frac{1}{2}\lambda M^2 - V) + \frac{1}{k_B}W \qquad (6\text{-}3)$$

$$T \leq \frac{1}{k_B}\lambda M^2(1-x)^2 - \frac{1}{k_B}(\frac{1}{2}\lambda M^2 + V) - \frac{1}{k_B}W \qquad (6\text{-}4)$$

$$(0 \leq x \leq 1, \ T \geq 0)$$

Now we draw the superconducting area of ineq.(6-3,6-4) by red full lines $A'H'B'$, and draw the superconducting area of ineq.(4-3,4-4) by black dotted lines $AHB$ as the same as Fig.1, they are compared in the Fig.2 below:

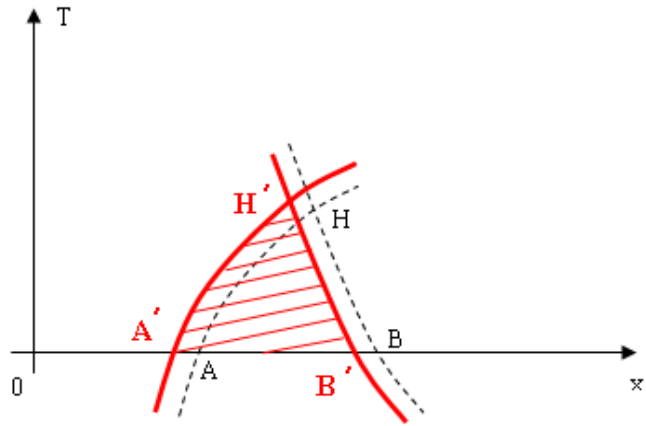

**Fig.2**: the influence of the electron-phonon interaction on real superconducting critical temperature.

We also remove the $(1-x)^2$ part of simultaneous inequations (6-3, 6-4), we get the relational expression as follow:

$$\frac{1}{2}\lambda M^2 + W \geq 3V + 3k_B T \qquad (6\text{-}5)$$

It is shown that the electron-phonon interaction energy $W$ can reduce the superconductivity threshold value of parent spontaneous magnetization.

In fig.2, the coordinate values of point $A'$, $B'$ $H'$ are as follows:

$$A' : (1 - \sqrt{1 - \frac{2(V-W)}{\lambda M^2}}\ ,\ 0)$$

$$B' : (1 - \sqrt{0.5 + \frac{V+W}{\lambda M^2}}\ ,\ 0)$$

$$H' : (1 - \sqrt{\frac{2}{3} + \frac{4W}{3\lambda M^2}}\ ,\ \frac{\lambda M^2}{6k_B} - \frac{V}{k_B} + \frac{W}{3k_B})$$

We can draw the following conclusions from the Fig.2:

(1) Relative to the area ABH neglected electron-phonon interaction, the real superconducting area $A'H'B'$ is left shift and upper shift appreciably. Namely It is shown that the enhancement of electron-phonon interaction is advantageous to the enhancement of $T_c$ value in weak doping area, but makes against the $T_c$ value in over-doping area.

(2) In Fig.2, the point H′ sits on the left side of the quantum critical point H, so the real best doping concentration x is less than 0.1835.

(3) If the electron-phonon interaction energy has a changed value $\Delta W$, the net variation value of the $T_c$ in weak doping area and over-doping area is $\frac{\Delta W}{k_B}$, but the $T_c$ variation value of the optimum doping point H′ is only $\frac{\Delta W}{3k_B}$. It is shown that the influence of phonon-electron interaction

alteration(such as isotope effect) on the optimum doping point is not obvious.

## 7. The pressure effect analysis

Under high pressure, we think three energy parameter values $\frac{1}{2}\lambda M^2$, $V$, $W$ become greater simultaneously, it seems very complex, but we can make use of the peak value $T_{c\max}$ of the point $H'$ to judge the comprehensive result.

Because,

$$T_{c\max} = \frac{\lambda M^2}{6k_B} - \frac{V}{k_B} + \frac{W}{3k_B}$$

So we use the $\dfrac{dT_{c\max}}{dP}$ to judge the influence of high pressure on superconductivity of cuprate superconductors,

$$\frac{dT_{c\max}}{dP} = \frac{d\left(\dfrac{\lambda M^2}{6k_B} + \dfrac{W}{3\,k_B}\right)}{dP} - \frac{d\left(\dfrac{V}{k_B}\right)}{dP}$$

The $\dfrac{dT_{c\max}}{dP}$ is a positive number or a negative number, it relates to the doped structure of cuprate superconductors.

Under high pressure, we think that the increasing of coulomb potential energy V between holes and electrons in p-type superconductors is less than the increasing of coulomb potential energy V between electrons and electrons in n-type superconductors. So we have reason to believe that as follows:

(1) To p-type superconductors,

$$\frac{d\left(\dfrac{\lambda M^2}{6k_B} + \dfrac{W}{3\,k_B}\right)}{dP} \geq \frac{d\left(\dfrac{V}{k_B}\right)}{dP},$$

Namely $\dfrac{dT_{c\max}}{dP} \geq 0$, so high pressure can usually raise the $T_{c\max}$ value before the lattice destruction.

(2) To n-type superconductors,

$$\frac{d\left(\frac{\lambda M^2}{6k_B}+\frac{W}{3k_B}\right)}{dP} \leq \frac{d\left(\frac{V}{k_B}\right)}{dP},$$

namely $\frac{dT_{cmax}}{dP} \leq 0$, so high pressure can not raise the $T_c$ value, on the contrary, the high pressure usually reduces the $T_c$ value in n-type superconductors.

## 8. Further study on cuprate superconducting mechanism

In this paper, we have thought that the doped holes of O2p orbital act as the attraction medium of cooper pairs, the doped electron holes of $Cu3d_{x^2-y^2}$ orbital have not the function of attraction medium.

But a doped hole of O2p orbital corresponds to a cooper pair namely two electron holes of adjacent $Cu3d_{x^2-y^2}$ orbital, so the doped holes of $Cu3d_{x^2-y^2}$ orbital is also important to superconductivity, and the ratio of holes number should follow $n_{x^2-y^2}/2n_\sigma \geq 1$, then the doped concentration x of O2p orbital electron holes should satisfy the extreme value condition $x \leq 1/3$. from the before-mentioned result $x \in (0, 0.2929)$, we know that it is in accordance with the condition $x < 1/3$.

## 9. The way to improve superconducting critical temperature

From the above-mentioned full text contents, we think that it is the way to improve superconducting critical temperature as following:

(1) To enhance the spontaneous magnetization of sublattice as much as possible;

(2) To enhance the electron-phonon interaction as much as possible;

(3) To have optimum doping of the O2p orbital electron holes;

(4) To decrease the coulomb repulsive potential between itinerant electrons as much as possible.

In practical applications, copper-rich treatment is a method of enhancing the spontaneous magnetization of sublattice; high pressure is also a possible method of improving superconducting critical temperature for p-type superconductors.

We should find more methods to improving the superconducting critical temperature.

## 10. Conclusion

In this paper, a series of important phenomena of cuprate superconductors are explained, we may summarize the following conclusions and viewpoints:

(1) spontaneous magnetization of sublattice is an effective parameter for studying and describing the magnetic fluctuation energy.

(2) two factors act on the microscopic pairing mechanism simultaneously, the main factor is the magnetic fluctuation, the secondary factor is the electron-phonon interaction, they are parallel relation.

(3) the doped hole of O2p orbital can be considered the attraction medium of cooper pairs, because its magnetic fluctuation energy has net attractive potential for ambient electrons.

(4) the doped holes of $Cu3d_{x^2-y^2}$ orbital is also important to superconductivity, they pay an important role of cooper pairs.

(5) full understanding of the dual characters of magnetic fluctuation and electron-phonon interaction is the key to study superconducting structure.

(6) using competitive energy relations to explore the superconductivity rule and energy criterion, we deduce a clear physical image of superconducting phase diagram.

(7) we get the way to improve superconducting critical temperature by physical parameter analysis.